# Simultaneous Assembly of van der Waals Heterostructures into Multiple Nanodevices

Enrique Burzurí,*[a] Mariano Vera-Hidalgo,[a] Emerson Giovanelli,[a] Julia Villalva,[a] Andres Castellanos-Gomez*[b] and Emilio M. Pérez*[a]

Van der Waals heterostructures (vdWH) are made of different two-dimensional (2D) layers stacked on top of each other, forming a single material with unique properties that differ from those of the individual 2D constituent layers, and that can be modulated through the interlayer interaction. These hetero-materials can be artificially made by mechanical stamping, solution processing or epitaxial growth. Alternatively, franckeite has been recently described as an example of a naturally-occurring vdWH that can be exfoliated down to nanometer thicknesses. Research on vdWHs has so far been limited to manually exfoliated and stamped individual devices. Here, a scalable and fast method to fabricate vdWH nanodevices from liquid phase exfoliated nanoflakes is reported. The transport and positioning of the flakes into localized submicrometer structures is achieved simultaneously in multiple devices *via* a dielectrophoretic process. The complex vdWH is preserved after dielectrophoresis and the properties of the resulting field-effect transistors are equivalent to those fabricated *via* mechanical exfoliation and stamping. The combination of liquid phase exfoliation and dielectrophoretic assembly is particularly suited for the study of vdWHs and applications where large-scale fabrication is required.

## Introduction

The exfoliation of bulk lamellar materials such as graphite, boron nitride, black phosphorous and other transition metal dichalcogenides down to the nanoscale led to the isolation of individual two-dimensional single layers and to the discovery of their unique electronic and optoelectronic properties[1,2]. Taking advantage of the van der Waals interlayer interaction in this family of materials, research in this field turned rapidly to the design and fabrication of tailored stacked heterostructures[3-5]. The main interest of such constructions lies in combining or modulating the properties of monolayers from different materials (doping, superconductivity, magnetism, light emission[4,6-10] among others) and exploiting their intrinsic thickness confinement to make new and ultrathin (thus potentially flexible) electronic devices.

The methods to fabricate such structures remain technically challenging. They mostly consist either in growing a single layer on top of another by epitaxy or stacking monolayers from two (or more) different materials by mechanical stamping[11]. Other techniques report a combined chemical co-exfoliation of 2D materials and solution processing to prepare heterostructure-based nanodevices.[12,13] However, in most cases, the building process is limited by crystallographic issues: van der Waals epitaxy does not require lattice matching but it is limited to a few materials, whereas in the mechanical stamping case, the correct alignment of the different crystal networks is difficult to achieve[14]. Moreover, the stamping processes are not exempt from possible interlayer contamination due to trapped molecules or adsorbates originating from precursors or molecules of the environment. As an alternative, we got interested in naturally occurring layered heterostructures from the sulfosalt family, and in particular in franckeite.[15-17] This crystalline sulfosalt is made of the alternate stacking of one tin(IV) sulfide ($SnS_2$)-like layer and four lead(II) sulfide (PbS)-like layers (Figure 1a). We recently evidenced,[18] concomitantly with another group,[19] that bulk franckeite can be exfoliated into nanoflakes that retain the heterostructure, both mechanically and through liquid phase exfoliation (LPE).[20-23] In addition, we demonstrated this material is an air-stable semiconductor with a very small bandgap (0.5-0.7 eV) and presents p-type doping, mostly due to substitutional Sb(III) in the structure, features that make ultrathin franckeite particularly interesting for the construction of miniaturized (opto)electronic devices.[24]

A challenging yet key step to improve toward device-making from exfoliated layers is the transfer and positioning of these lamellar heterostructures into submicrometer-scale localized areas of electronic devices, like the drain-source electrodes in a transistor. This problem has been partially addressed for prototyping, using mechanical exfoliation and stamping techniques,[25] which unfortunately are not suitable for nanometer scale flakes and for large-scale industrial applications.

Here we report the controlled transport and assembly of LPE colloidal two-dimensional franckeite heterostructures (PbS-$SnS_2$) simultaneously into multiple localized submicrometer-spaced electrodes. This high selectivity in position within large-scale areas is possible thanks to the combination of the LPE of the bulk material with the application of a dielectrophoretic field to the solution. Dielectrophoresis (DEP) consists in the directed motion of dielectric and polarizable particles properly dispersed within a liquid medium under the influence of an electric field gradient. Such a gradient can be generated by the application of an alternating voltage between two electrodes

[a.] *IMDEA Nanoscience, Ciudad Universitaria de Cantoblanco, c\Faraday 9, 28049 Madrid, Spain.* Email: enrique.burzuri@imdea.org, emilio.perez@imdea.org
[b.] *Materials Science Factory. Instituto de Ciencia de Materiales de Madrid (ICMM-CSIC), c\Sor Juana Inés de la Cruz 3, 28049 Madrid, Spain.* Email: andres.castellanos@csic.es

and has been used before to trap low-dimensional objects like nanoparticles[26-28], carbon nanotubes[29-31], nanoribbons[32] and nanowires[33]. We show that the lamellar structure of franckeite is preserved after DEP, something *a priori* not straightforward for heterostructures formed from individual layers with different polarizabilities. The resulting devices are arrays of franckeite-based field-effect transistors (FET), connected in parallel, of characteristics comparable to those obtained *via* mechanical stamping of a flake[18].

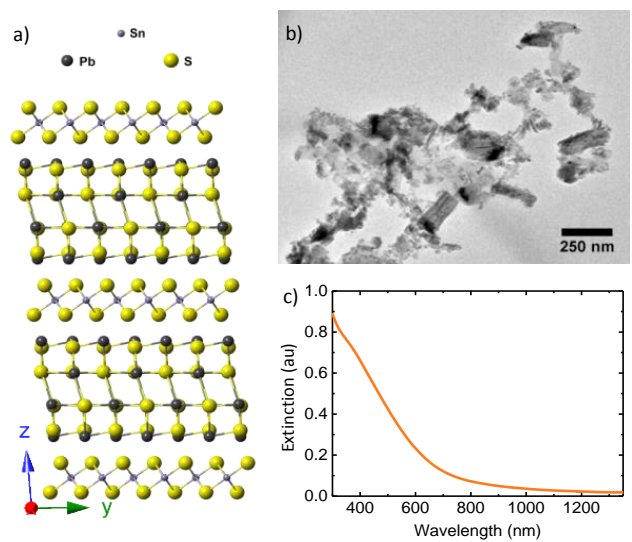

Fig. 1 Franckeite (PbS-SnS$_2$) and its liquid-phase exfoliation in *i*PrOH. a) Franckeite simplified crystal structure (adapted from Makovicky *et al.*[17] and with Shannon's crystal ionic radii[34]). The stoichiometry of our samples is Pb 25.6 Sn 9.0 Fe 0.7 Sb 10.0 S 53.7 Ag 1.0 as determined from XPS. b) TEM image of typical nanoflakes from the colloidal suspension obtained after exfoliation of a 1 mg.mL$^{-1}$ franckeite dispersion in *i*PrOH and subsequent centrifugation. c) UV-Vis-NIR spectrum of franckeite colloidal suspension.

## Results and discussion

LPE of franckeite proved to be efficient in producing colloidal flakes having few-unit-cell thicknesses[18, 19] in various neat polar solvents such as *N*-methylpyrrolidone (NMP), methanol, isopropanol (*i*PrOH), water and *i*PrOH/water mixtures. The exfoliation proceeds smoothly without the help of any additional surfactant species. These solvents possess a surface tension high enough to overcome the van der Waals interaction between the heterostructured layers; additionally, they have dielectric and coordination properties that ensure the proper colloidal dispersion of the resulting nanoflakes. As a consequence, unlike other wet-phase exfoliating techniques (such as ion intercalation or surfactant-assisted exfoliation), this method leaves as few insulating organic residues (solvent molecules) as possible on the layer surface and does not necessitate any washing step, which is particularly convenient for device fabrication.

Within the set of possible solvents, and with the aim of fabricating devices *via* DEP, a first critical issue is the solvent polarizability compared to that of the LPE franckeite flakes. The DEP force will drive the nanoflakes toward the gap between the electrodes only if the colloidal particles are more polarizable than the surrounding medium (see the Supporting Information). Pure *i*PrOH shows relatively weak polarizability (6.98 Å$^3$)[35] and exhibits a boiling point of 82 ºC, which makes it non-volatile enough to perform DEP within a drop-cast suspension, but also easy enough to evaporate to complete the device and connect the nanoflakes amongst themselves and with the electrodes. Finally, our previous experiments on LPE franckeite showed that exfoliation in *i*PrOH leads to a flake lateral size distribution centered between 100 nm and 200 nm, which is suited for the nanoplatelets to fill the gap (a few hundreds of nanometers wide) between the electrode pairs. See supporting information for discussion of other possible solvents.

The suspension of LPE franckeite in *i*PrOH was thus obtained via a two-step process: sonication of a 1 mg.mL$^{-1}$ dispersion of franckeite powder in *i*PrOH, followed by centrifugation to discard non-exfoliated material (see the Supporting Information). The light orange colloid presents the expected typical features of suspended franckeite nanoflakes in *i*PrOH,[18] as demonstrated by TEM: 100-200 nm sized flakes, Figure 1b; UV-Vis-NIR: continuous extinction starting from the near-infrared region, Figure 1c; and Raman spectroscopy (discussed below).

The target devices for the heterostructures consist of a set of submicrometer-spaced Au electrodes fabricated by laser mask-less optical lithography and subsequent Au thermal evaporation and lift-off. Tens of these electrodes are connected to a pair of Au pads that serve as common electrical contacts. In addition, the devices lie on a heavily doped Si substrate, coated with a thin (300 nm) silicon oxide layer that is used as a back-gate electrode (see the Supporting Information for additional details and images of the full device). Franckeite-based FETs are fabricated by DEP assembly of the nanoflakes between these electrodes, as schematically shown in Figure 2a. An *i*PrOH micro-droplet containing suspended franckeite nanoflakes is drop-casted onto the device. Thereafter a DEP AC voltage ($V_{AC}$ = 10 V, ν = 1 MHz) between the electrodes is applied for 10 minutes to promote the mobility of the flakes into the inter-electrode space. Figure 2b shows a color map of the square of the electrical field distribution in the device calculated by finite element analysis for a bias difference of 10 V between the electrodes. The tip-like design of the electrodes allows focusing strong electrical fields in a small confined area, reaching 10$^7$ V/m. Importantly for DEP, the electrical field rapidly decays by orders of magnitude when moving only a few micrometers apart from the gap between the electrodes, as seen in the inset of Figure 2c. The electrical gradient generated is therefore large and directed towards the inter-electrode space as seen in Figure 2c. Note that since the electrical field is maximized in the gap, the gradient is zero and therefore the flakes are trapped once they reach that area (see the Supporting Information for more details). After the DEP assembly, the device is thoroughly rinsed with *i*PrOH and dried with nitrogen gas to remove excess material.

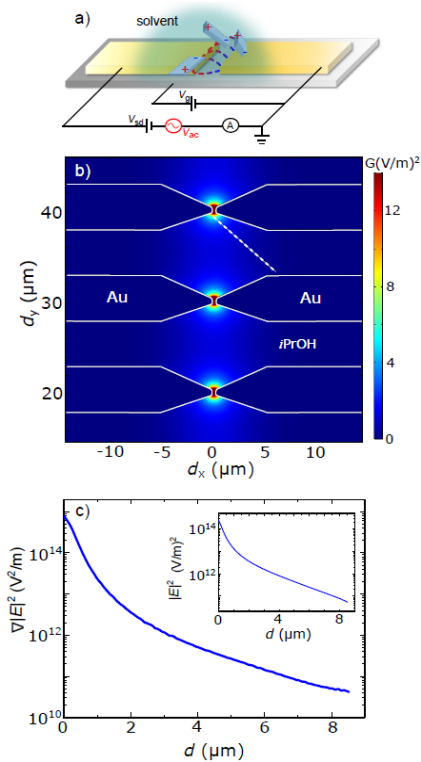

Fig. 2 DEP assembly of layered heterostructures. a) Schematic representation of the DEP of colloidal heterostructures between metallic electrodes. b) Squared electrical field distribution ($E^2$) color map simulated *via* finite element analysis at a bias $V$ =10 V. The electrical field is maximized in the gap between electrodes. The gradient of the electrical field ($\nabla|E|^2$) points to and increases towards that area. The electrodes' edges are highlighted in white for clarity. c) Gradient of $|E|^2$ taken along the dotted line in (b). The inset shows the corresponding $|E|^2$.

Figure 3a shows a scanning electron microscopy (SEM) image of a representative multi-electrode device after DEP assembly. Additional images can be found in the Supporting Information. Most of the deposited material appears located within the gaps between the electrodes and decorating their edges where the electrical field and the electrical gradient are more intense (see also Supporting Information). We find that around an 85% of the electrodes in the device are connected by flakes. This yield can be improved by fine-tuning the DEP parameters. A closer inspection into a pair of electrodes (Figure 3b) reveals tens of nano-crystallites filling the inter-electrode gap while the surrounding surface remains relatively clean preventing inter-device short-circuits. The aspect ratio of the crystallites appears homogeneous with a flake lateral size between 100-200 nm as observed in the TEM images. The stacking of the material between the electrodes is further analyzed in the atomic force microscopy (AFM) image in Figure 3c. The image and the height profile taken across two gaps show accumulation of material 200-400 nm high and 1 μm wide, corresponding to approximately the width of the electrode tip and four times its height. Additional images can be seen in the Supporting Information. DEP is therefore a fast and efficient method to simultaneously transport and assemble lamellar heterostructures into complex nanometer-scale device structures. Note that the simultaneous transfer of numerous nanometer-scale flakes and the high positional selectivity on the device is not achievable with conventional mechanical techniques, like stamping, nor by simply drop-casting liquid phase exfoliated colloidal suspensions as shown in the Supporting Information.

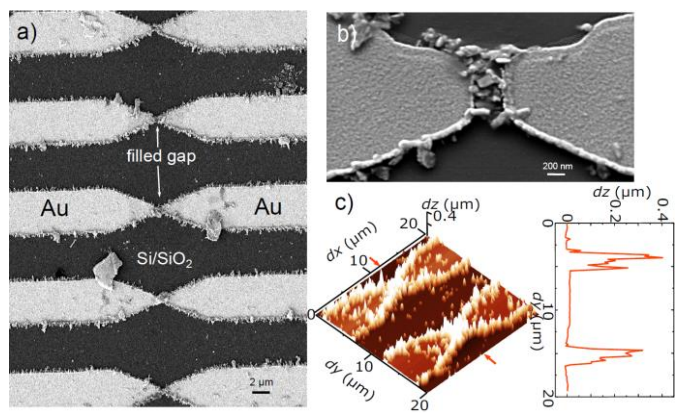

Fig. 3 Imaging of the devices after DEP assembly. a) Scanning electron microscopy (SEM) image of a multi-electrode device. DEP transports and confines the nano-flakes into the inter-electrode area where the electric field is more intense. b) Closer image of a single pair of electrodes. Tens of crystallites are packed between the electrodes while the surrounding surface is relatively clean. c) Atomic force microscopy (AFM) image of two pairs of electrodes and the corresponding height profile taken across the gaps (marked by the arrows). The image shows accumulation of franckeite up to 400 nm high and 1 μm wide (the width of the electrode tip).

The transfer characteristic measured in the same device at a fixed source-drain voltage $V_{sd}$ = 10 V is shown in Figure 4a. The measured current $I$ reaches the 0.1 μA range at negative $V_g$ and drops around an order of magnitude at positive $V_g$. This behavior, indicative of hole-carrier depletion in the material, is characteristic of a p-doped semiconductor, as previously reported for FETs based on mechanically exfoliated franckeite.[18, 19, 24] The electronic noise is most likely electromechanical noise due to the flake-to-flake mechanical contact that can be modified with the gate electric field and the current flow across the device. A given device is rather stable, even in ambient conditions and operating at high voltages.

Figure 4b shows the current-voltage ($I$-$V_{sd}$) characteristics measured at zero gate voltage. The non-linear response of the current to a bias voltage is indicative of the formation of Schottky barriers between the Au electrodes and the franckeite nanoflakes or possibly tunneling between the flakes. As a control experiment, we have compared these results with other devices prepared by drop-casting LPE franckeite in the absence of a DEP field. The images taken under the optical microscope show a random distribution of franckeite flakes over the device (see supporting information). In addition the current levels are orders of magnitude lower than in devices prepared *via* DEP assembly. DEP assembly of lamellar heterostructures, franckeite, can thus produce FET devices with electronic properties equivalent to those produced *via* mechanical exfoliation and stamping.

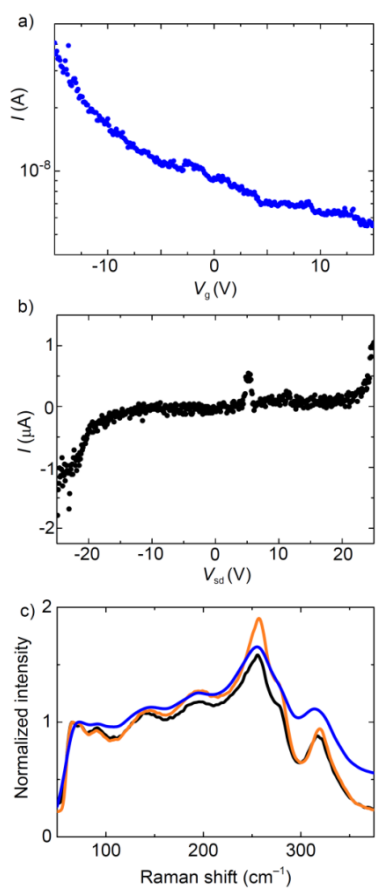

Fig. 4 Electronic and spectroscopic characterization after DEP assembly. a) Transfer characteristic measured at a fixed $V_{sd}$ = 10 V. The drop in current at positive gates voltage ($V_g$) is fingerprint of the hole-carrier depletion in a p-doped semiconductor, as observed in mechanically-exfoliated franckeite devices. b) Current-voltage ($I$-$V_{sd}$) characteristic measured at zero gate voltage. The non-linear response in bias is indicative of the formation of Schottky barriers between the Au electrodes and the franckeite flakes c) Raman spectra ($\lambda_{exc}$ = 532 nm) of bulk (orange), colloidal franckeite (black) and franckeite after DEP in the device (blue).

To further explore the structural integrity of the heterostructures after the DEP assembly and the electron transport measurements, Raman spectroscopy of the flakes placed between electrodes is compared with bulk and colloidal franckeite. The normalized spectra are shown in Figure 4c. The five main bands characteristic of franckeite[24] (70, 143, 195, 256 and 318 cm$^{-1}$) are observed in the three cases with minor variations, mainly broadening of the signals, most likely due to the presence of the gold electrodes. See supporting information for the identification of the Raman bands. Franckeite is therefore sufficiently stable to withstand strong DEP fields and bias voltages. This result is a priori not straightforward since the different polarizabilities of the heterogeneous component layers in the van der Waals heterostructure could create strains in the material that might have led to disassembly.

## Conclusions

In summary, we report a scalable method to fabricate nanodevices based on van der Waals heterostructures. We combine LPE of natural lamellar heterostructures with DEP transport and assembly to simultaneously fabricate multiple franckeite-based FETs. We show that this technique shows both high selectivity in the size of the flake, in the submicrometer range, and high site selectivity in their positioning in complex devices. Raman spectroscopy and electron transport measurements show that the properties of the franckeite heterostructure are preserved after the assembly in the device, and that the electronic properties of the resulting devices are comparable to devices fabricated with mechanically exfoliated franckeite flakes. Our results open the door to the study and scaling in large area devices of large families of natural layered heterostructures, like the sulfosalts, but also artificial heterostructures like misfit layer compounds and ferecrystals[36] and custom-made heterostructures obtained by solution processing.[12, 13]

## Conflicts of interest

There are no conflicts to declare.

## Acknowledgements

Funding from the European Union (ERC-Starting Grant: 307609 (EMP), MINECO (Grants: CTQ2014-60541-P (EMP) and the Comunidad de Madrid (Grant: MAD2D-CM program S2013/MIT-3007 (EMP)) is gratefully acknowledged. E.B. thanks the AMAROUT-II fellowship (Marie Curie Action, FP7-PEOPLE-2011-COFUND) and MSCA-IF 746579. IMDEA Nanociencia acknowledges support from the 'Severo Ochoa' Programme for Centres of Excellence in R&D (MINECO, Grant SEV-2016-0686). AC-G acknowledges funding from the European Commission under the Graphene Flagship, contract CNECTICT-604391. We thank the National Centre for Electron Microscopy (ICTS-CNME, Universidad Complutense) for electron microscopy facilities.

# Supporting Information

**Simultaneous Assembly of van der Waals Heterostructures into Multiple Nanodevices**

*Enrique Burzurí\*, Mariano Vera-Hidalgo, Emerson Giovanelli, Julia Villalva, Andres Castellanos-Gomez\* and Emilio M. Pérez\**

This Supporting Information is divided in the following sections: (1) Experimental procedures, (2) fabrication details and schematics of the multi-electrode devices, (3) dielectrophoresis mechanism and possible solvents, (4) finite elements analysis of the electrical field distribution, (5) additional SEM images of the devices, (6) additional AFM characterization of the device, (7) electric characterization of additional devices, (8) a comparative with dropcasted devices and (9) Supplementary table with the interpretation of the Raman spectra.

### (1) Experimental procedures

*Preparation of franckeite colloidal suspension.* Chips from natural franckeite mineral (San José mine, Oruro, Bolivia) were ground in an agate mortar until a fine black powder was obtained. Franckeite powder (10 mg) was dispersed in *i*PrOH (10 mL) in a 20-mL glass vial. The dispersion was subjected to ultrasound irradiation for 1 h in an ultrasonic bath (Fisher Scientific FB 15051; 37 kHz, 280 W, ultrasonic peak max. 320 W, standard sine-wave modulation) connected to a cooling system maintaining the water bath temperature at 20 °C. The resulting black suspension was centrifuged at 990 *g* and 20 °C for 30 min (Beckman Coulter Allegra X-15R, FX6100 rotor, radius 9.8 cm); it separated into a black sediment and an orange supernatant, which was carefully isolated from the solid. The corresponding franckeite suspension remained colloidally stable for 48 h to 72 h, after which it progressively deposited. Nonetheless, the franckeite flakes could easily be redispersed by 1-2 min bath sonication.

*Transmission Electron Microscopy (TEM).* The colloidal suspension was drop-casted onto a 200 square mesh copper grid covered with a carbon film. After a few minutes, the excess solvent was removed and the grid was left drying in the air at room temperature. The procedure was repeated 5 times and the grid was finally dried under vacuum for 48 h. The observation was performed using a JEOL JEM 2100 microscope operated at 200 kV.

*UV-Vis-NIR spectroscopy.* As-prepared colloidal suspension was transferred to a quartz cuvette and its extinction spectrum (sum of the absorption and scattering spectra) was measured using a Cary 5000 spectrophotometer from Agilent Technologies.

*Raman spectroscopy.* Bulk franckeite powder (pressed onto a glass slide), liquid-phase-exfoliated franckeite (drop-cast on a glass slide and dried at 40 °C several times) and on-device franckeite (after DEP) were characterized using a Bruker Senterra confocal Raman microscope (Bruker Optik, Ettlingen, Germany; objective NA 0.75, 50×; laser excitation: 532 nm, 0.2 mW). The spectra result from the average of 10 measurements acquired from different regions over the whole samples.

*Device fabrication.* The multi-electrode devices are fabricated *via* laser mask-less optical lithography and thermal evaporation of Cr/Au (5/70 nm). A lift-off process in acetone/*i*PrOH/deionized water removes the excess metallic material. The devices are fabricated on a highly-doped silicon substrate capped with a 300 nm thick insulating $SiO_2$ layer. This substrate is used as common back-gate electrode. Additional details and the final device are shown in the Supporting Information.

*Atomic Force Microscopy.* The AFM images were acquired in intermittent (tapping) mode and under ambient conditions by using a NT-MDT NTEGRA PRIMA station equipped with a SF005$AU007NTF head and NT-MDT NSG01 silicon cantilevers with typical spring constant and resonant frequency of 5.2 $Nm^{-1}$ and 144 kHz respectively.

*Electron Transport Measurements.* The current-voltage curves and transfer characteristics were obtained in ambient conditions in the chamber of an electrical probe station equipped with a Keithley 2450 digital source-meter unit.

## (2) Fabrication details and schematics of the multi-electrode devices

The multi-electrode devices (see Figure S1) are fabricated *via* laser mask-less optical lithography and thermal evaporation of Cr/Au (5/70 nm) electrodes. The finger-shaped electrodes are connected to common Au pads that allow performing simultaneous dielectrophoresis to all the devices. A lift-off process in acetone/iPrOH/deionized water removes the photoresist and the excess metallic material. The devices are fabricated on a highly-doped silicon substrate coated with a thin insulating $SiO_2$ layer. This substrate is used as common back-gate electrode. A scheme and a scanning electron microscopy (SEM) image of final device are shown in Figure S1.

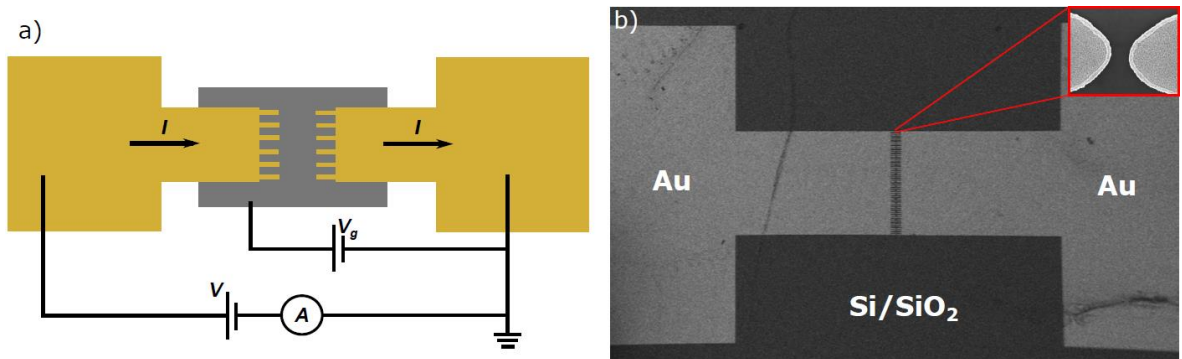

**Figure S1.** (a) Schematics and (b) scanning electron microscopy (SEM) image of the full multi-electrode device.

## (3) Dielectrophoresis mechanism and possible solvents

The dielectrophoretic (DEP) force $F_{DEP}$ exerted on oblate ellipsoidal particles with a large aspect ratio (which best approximates exfoliated layers) can be written as follows:

$$F_{DEP} \propto V_p \, \varepsilon_m \, Re\left[\frac{\varepsilon_p^* - \varepsilon_m^*}{\varepsilon_m^*}\right] \nabla |E|^2$$

where $V_p$ is the volume of the particles; $\varepsilon_p^*$ and $\varepsilon_m^*$ are respectively the complex permittivities of the particles and the suspension medium; and $E$ is the non-uniform electric field. The DEP force is thus proportional to the nanoflake volume, aligned with the electric field gradient and its orientation depends on the sign of the real part of the Clausius-

Mossotti factor $\left[(\varepsilon_p^* - \varepsilon_m^*)/\varepsilon_m^*\right]$. The Clausius-Mossotti factor is in turn related with the polarizability of the particles and solvent.

The selection of *i*PrOH as solvent is made on the basis of its weaker polarizability (6.98 Å$^3$[1]) compared to the constituents of franckeite (~10 Å$^3$) [2,3], necessary for dielectrophoresis, and a boiling point (82 ºC) that makes *i*PrOH non-volatile enough to perform DEP but easy enough to evaporate. In addition, LPE in *i*PrOH produces colloidal flakes having few-unit-cell thicknesses as we explain in the main text.

These are general conditions that any solvent need to meet in order to be suitable for LPE+DEP. Liquid phase exfoliation of franckeite has been achieved in *i*PrOH, different water/ *i*PrOH mixtures and methanol as reported in [4], as well as in acetonitrile (data not reported). This could probably be expanded to alcohols in general although it remains to be proved.

Once the LPE is achieved, DEP of the colloids could be performed with solvents presenting a polarizability lower than 10 Å$^3$ and a boiling point between 60 ºC and 90 ºC. Some solvents meeting those requirements (see [1]) are 2,2,2-trifluoroethanol (5.20 Å$^3$; 78°C), 2-propenenitrile (6.24 Å$^3$; 77°C), acetonitrile (4.44 Å$^3$; 82°C), ethanol (5.13 Å$^3$; 78°C) and methanol (3.26 Å$^3$; 65°C). Other solvents like acetone (6.47 Å$^3$; 56°C), ethyl formate (7.09 Å$^3$; 54°C) and methyl acetate (7.00 Å$^3$; 57°C) could be used although the lower boiling point could present problems due to a fast evaporation.

Some of these solvents are not necessarily good for LPE and a single solvent for the whole process would be desirable. Taking this into account, acetonitrile, *i*PrOH, methanol and ethanol are good candidates for combined LPE and DEP that in addition are commonly used solvents.

## (4) Finite elements analysis of the electrical field distribution in the device

The electrical field distribution in the device surface is calculated by using a finite elements analysis software. A real-scale pattern (2D in-plane for simplicity) of the device is imported in the software where the electrostatic parameters (voltage potential, charge conservation, dielectric constants) of the boundaries are assigned. The pattern is thereafter divided in a software-optimized polygonal mesh. The computed electrical field is plotted in Figure 2 of the main manuscript.

Figure S2(a,c) shows the $E^2$ and $\nabla|E|^2$ profiles taken along the $y$ axis and crossing the gap area between the electrodes (yellow line in the inset of Figure S2d). The electrical field is maximum within the gap between the electrodes. The gradient is directed towards the gap where it becomes zero. The flakes are therefore trapped once they reach the inter-electrode space. This allows the controlled accumulation of material by graduating the time of dielectrophoresis and concentration of flakes in the liquid phase exfoliation

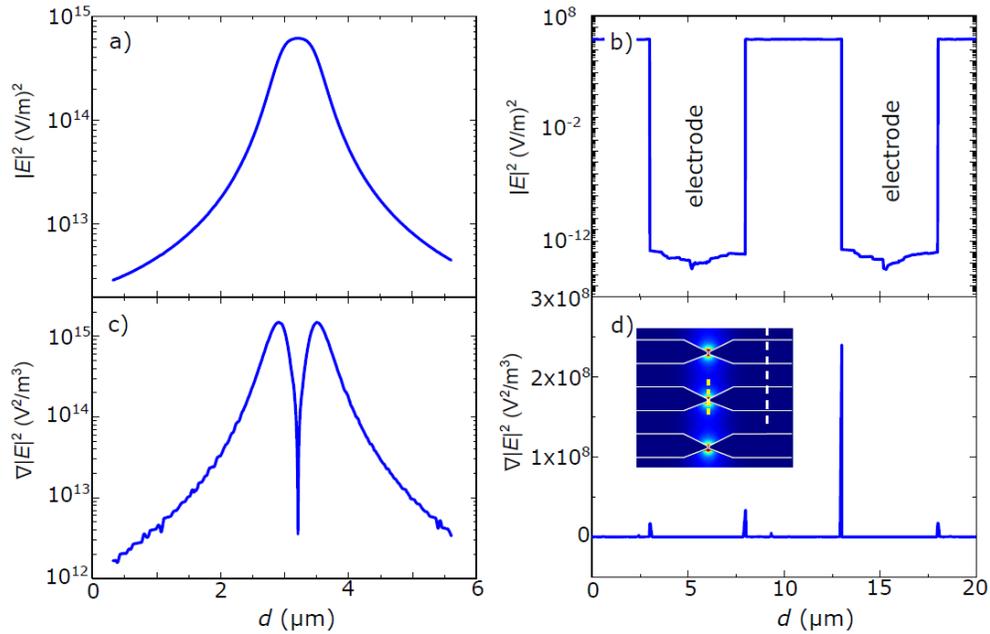

**Figure S2**. (a) Squared electrical field ($E^2$) and (c) $\nabla|E|^2$ across the gap in the $y$-axis along the yellow line in (d) inset. (b) Squared electrical field ($E^2$) and (d) $\nabla|E|^2$ in the $y$-axis in the "inter-finger" area marked by the white line in the inset. The inset shows the $E^2$ distribution in the device. The electrodes' edges are highlighted in solid white as a guideline to the eye.

Figure S2(b,d) shows the $E^2$ and $\nabla|E|^2$ profiles taken in the area far from the gap and in between the Au "finger-like" electrodes (white line in the inset of Figure S2d). The in-plane electrical field is around four orders of magnitude smaller than in the area around the gap. This makes $|E|^2$ eight orders of magnitude smaller. However, the sharp transition between the dielectric and the electrode area creates a strong electrical gradient at the edge of the electrodes. The flakes are therefore directed towards the edge of the electrode as has been observed experimentally.

### (5) Additional SEM images of the franckeite-based devices

Figure S3 shows additional SEM images of pairs of electrodes after dielectrophoresis. The franckeite flakes are mainly packed in the inter-electrode space, where the dielectrophoretic force is directed, and along the electrode edges as shown in the main text and discussed before. The surrounding substrate is clean of material.

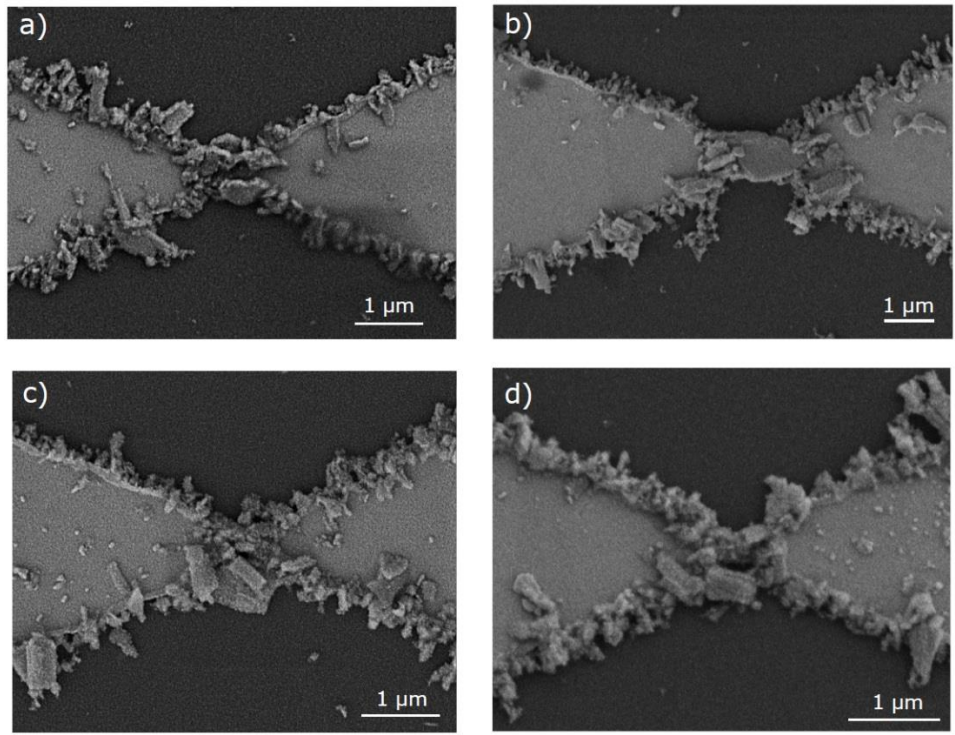

**Figure S3.** Scanning electron microscopy (SEM) images taken in different electrode-pairs of different devices. The material is mainly found in the inter-electrode space and the electrode edges.

**(6) Additional AFM images of the franckeite-based devices**

Figure S4 shows atomic force microscopy (AFM) images of the bare electrodes before DEP and the liquid-phase exfoliated franckeite flakes drop-casted on a Au substrate. The height profile of the flakes fluctuates between 50 nm and 150 nm, possibly due to the stacking of several flakes.

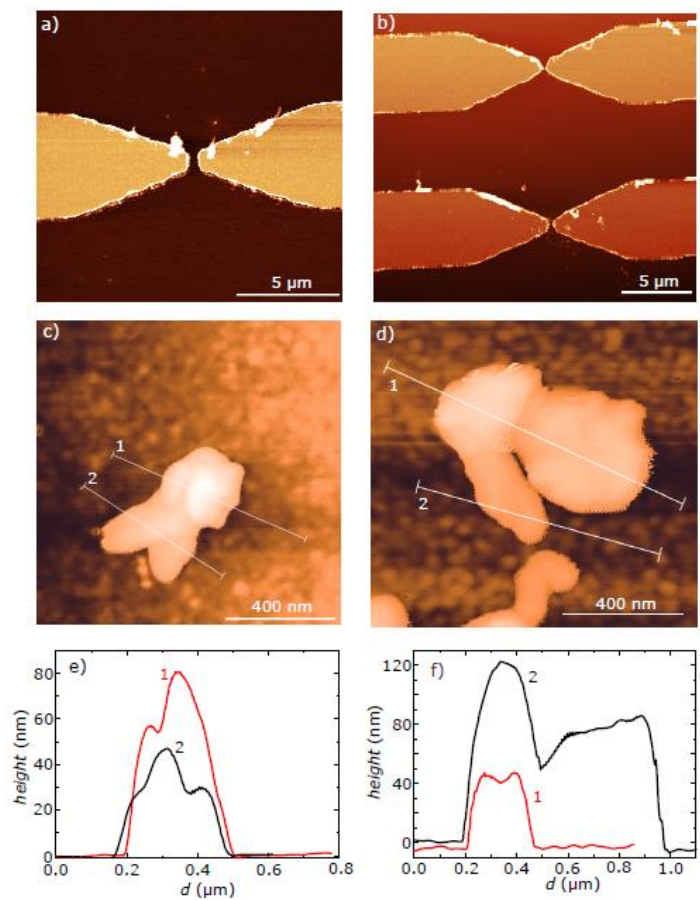

**Figure S4.** (a), (b) Atomic force microscopy (AFM) images of the empty electrodes. (c), (d) AFM images and (e), (f) height profiles of franckeite flakes deposited on a Au substrate.

After dielectrophoresis, the flakes gather mainly in the gaps creating stacks that reach around 1 μm wide and around 400 nm high as seen in different devices in Figure S5. These values compared with the average dimensions of flakes indicate that three to five flakes may be stacked in height within the gap.

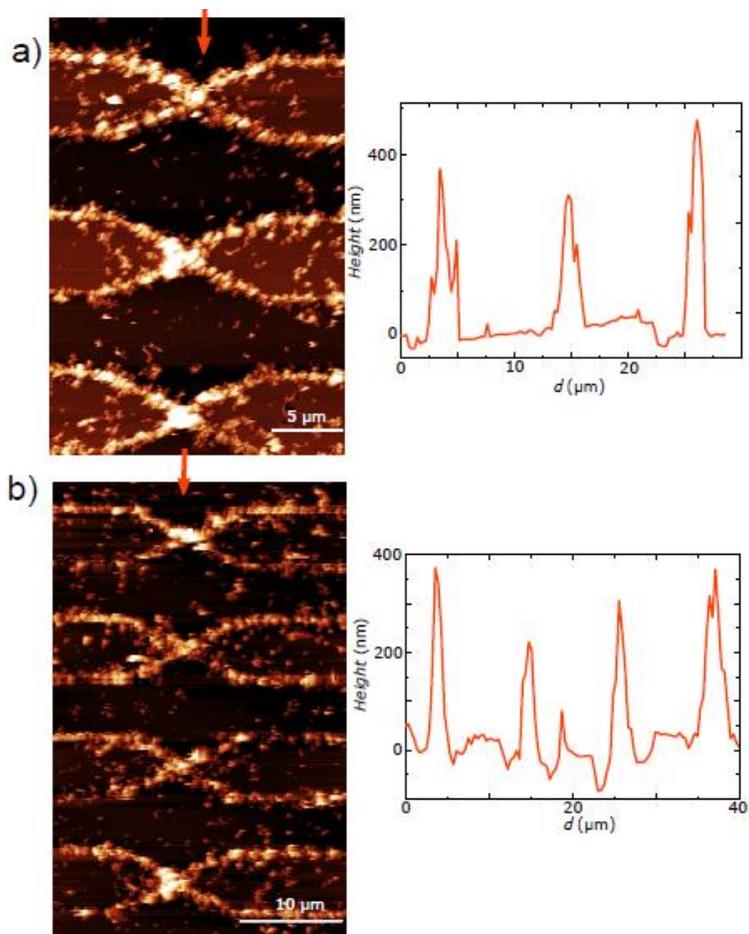

**Figure S5.** Atomic force microscopy (AFM) images and corresponding height profiles across the gaps taken in different devices.

### (7) Electrical characterization of additional devices

Figure S6 shows the current ($I$) - voltage ($V$) curves and the corresponding transfer characteristic measured in an additional franckeite-based multi-electrode device. The negative slope of the transfer characteristics is indicative of a p-doped semiconducting behavior as reported for bulk franckeite. Besides, the non-linear *I-V* curves are indicative of the formation of Schottky barriers between the semiconducting franckeite and the metallic Au electrodes.

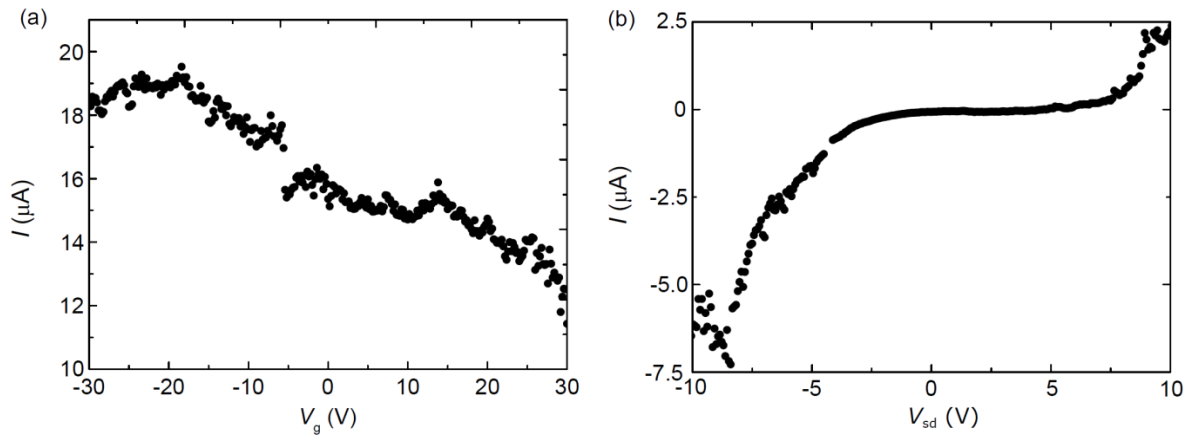

**Figure S6**. a) Transfer characteristic and b) current-voltage characteristic of an additional multi-electrode franckeite-based device. The transfer characteristic points to a p-doped semiconductor. The asymmetric *I-V* curve indicates the formation of Schottky barriers between the metallic electrodes and the semiconducting franckeite.

### (8) Devices prepared by drop-casting

Figure S7a shows an optical microscope image of a device prepared by DEP. The franckeite is clearly concentrated around the electrodes tips. The DEP conditions are set assemble more material and therefore facilitate visualization under the optical microscope. In contrast, Figure S7b shows a multi-electrode device prepared by drop-casting of a liquid-phase exfoliated franckeite droplet in the absence of a dielectrophoretic field. The franckeite appears evenly distributed over all the surface and no significant accumulation between the electrodes is observed.

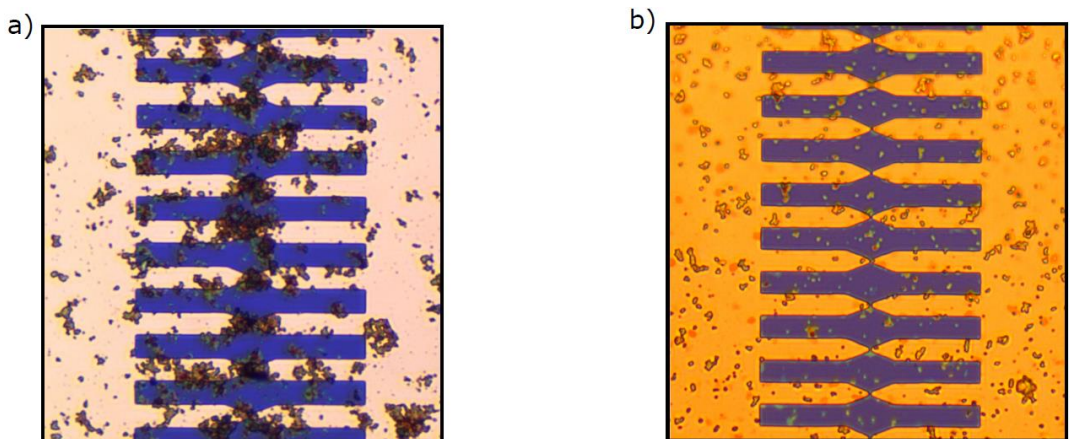

**Figure S7**. Optical microscope images of multi-electrode devices made by (a) dielectrophoresis and (b) drop-casting in the absence of a dielectrophoretic field. The franckeite is clearly concentrated around the electrodes tips in a) whereas is evenly dispersed in b). The DEP conditions in a) are set to assemble more material than in the main manuscript to facilitate the visualization under the optical microscope.

### (9) Supplementary tables

Supplementary Table 1 shows the labelling and interpretation of the different Raman modes of franckeite as reported in the Supporting Information of Ref [4].

**Supplementary Table 1. Interpretation of the Raman spectra of franckeite [4]**

| Raman shift (cm$^{-1}$) | Phonon mode attribution | Compound |
|---|---|---|
| 70 | Acoustic | PbS |
| 143 | 2$^{nd}$ order effect | SnS$_2$ |
|  | Transverse acoustic and transverse optical | PbS |
| 195 | Longitudinal optical | PbS |
|  | E$_g$ | SnS$_2$ |
| 256 | Combination | PbS+SnS$_2$ |
| 318 | A$_{1g}$ | SnS$_2$ |